\documentclass[12pt,a4paper,notoc]{JHEP}
\usepackage{epsfig}
\usepackage{amsfonts}
\usepackage{amssymb}

\newcommand{\be}{\begin{equation}}
\newcommand{\ee}{\end{equation}}
\newcommand{\bd}{\begin{displaymath}}
\newcommand{\ed}{\end{displaymath}}
\newcommand{\bea}{\begin{eqnarray}}
\newcommand{\eea}{\end{eqnarray}}
\newcommand{\non}{\nonumber}
\newcommand{\hlf}{\frac{1}{2}}

\title{Non--trivial flat connections on the 3--torus II\\ \large{The exceptional groups $F_4$ and $E_{6,7,8}$}}

\author{Arjan Keurentjes\\Instituut-Lorentz for theoretical physics, Universiteit Leiden \\ P.O. Box 9506, NL-2300 RA Leiden, 
The Netherlands\\E-mail \email{arjan@lorentz.leidenuniv.nl}}

\abstract{We continue the construction of non-trivial vacua for gauge theories on the 3--torus. Application of constructions based on twist in $SU(N)$ with $N > 2$ produce more extra vacua in theories with exceptional groups. We calculate the relevant unbroken subgroups, and their contribution to the Witten index. We show that the extra vacua we find in the exceptional groups are sufficient to solve the Witten index problem for these groups.}

\preprint{INLO-PUB-5/99}

\begin{document}
\section{Introduction}

The vacuum equations for Yang-Mills theories on the 3--torus with periodic boundary conditions $F_{\mu \nu} =0$ always allow for the trivial solution $A_{\mu} = c^{a}_{\mu} H_a$ with $c^a_{\mu}$ a constant and $H_a$ group generators in a maximal abelian algebra of the group (Cartan subalgebra (CSA)). For $SU(N)$ and $Sp(N)$-theories, all solutions belong to the trivial class, but for other gauge groups this is not the case. In \cite{Witnew, Keur1} it was shown that other vacua, not belonging to the trivial class, exist for theories with $G_2$ and $SO(N)$ ($N > 7$) gauge group. In a previous article \cite{Keur2} we proposed a systematic approach to constructing non-trivial vacua in gauge theories on the 3--torus, and applied the method to reproduce the results of \cite{Witnew, Keur1}, and embed these results in the exceptional groups $F_4$ and $E_{6,7,8}$. In this article  we give the details for the extra vacua in theories with exceptional groups, as was announced in \cite{Keur2} (see table 1 and 2).

The non-trivial vacua of a gauge theory on the 3--torus are of relevance for the problem of computing ${\rm Tr} (-1)^F$ in supersymmetric gauge theories \cite{IWit}. As ${\rm Tr}(-1)^F$ is an index, assuming no singularities as a function of the volume, this Witten index should be equal in the infinite volume. In the infinite volume the theory has a chiral $U(1)$ symmetry, broken to a discrete $\mathbf{Z}_{2h}$ symmetry by instantons, where $h$ is the dual Coxeter number for the gauge group $G$. Assuming gluino condensation, the symmetry is broken to a $\mathbf{Z}_2$, leaving $h$ different vacua \cite{IWit, IDyn}. In the finite volume version on $T^3 \times R$, the moduli space of flat connections might consist of several components. In each component of the moduli space, the gauge group is partially broken. Performing a Born-Oppenheimer approximation each component of the moduli space, which represents a particular class of solutions, contributes $r'+1$ to ${\rm Tr} (-1)^F$, where $r'$ is the rank of the unbroken gauge group in the relevant component. Equating ${\rm Tr} (-1)^F$ for the infinite volume to the result for the finite volume one should find \cite{Witnew} 
\be \label{Cox}
h = \sum_i (r_i + 1)
\ee
where the sum runs over disconnected components of moduli space, i.e. different classes of solutions, and $r_i$ is the rank of the unbroken gauge group in each of these components. This equality is satisfied for theories with $SU(N)$ and $Sp(N)$-gauge groups \cite{IWit}, where the moduli-space consists of only one component. For the theories with $SO(N)$ groups a second component of the moduli-space was found in \cite{Witnew}, and for $G_2$ the moduli space also consists of two components \cite{Keur1}. In both cases the identity (\ref{Cox}) is satisfied. For the remaining exceptional groups, a first step towards solving the problem was presented in \cite{Keur2}, by showing that the results of \cite{Witnew, Keur1} can be embedded in these exceptional groups. 

{\bf Note:} After the publication of the first part of this article \cite{Keur2} in which the results of this paper were announced, another paper on the same subject appeared \cite{Kac}. The results in this paper agree with ours. 

\section{Method of construction}

We will shortly review our method. Instead of giving an explicit expression for the gauge potential, we construct a set of commuting holonomies in a simply connected representation of the gauge group $G$ (in fact, since they commute in a simply-connected representation of the gauge group, they will commute in any representation). A simple theorem \cite{Keur1} then guarantees a corresponding flat connection. To construct commuting elements of a given group, we study a $PG'_N$-subgroup, whose universal covering $\widetilde{PG'}_N$ is a product of $N$ factors $\tilde{G}'$. The global structure of $PG'_N$ is not that of a product group, since there is a diagonal central subgroup divided out. This allows us to impose so-called multi-twisted boundary conditions \cite{Cohen}. As an example we consider a subgroup $PG_2'$ (with universal covering $\tilde{G}'^2$). We assume that $\tilde{G}'$ allows a non-trivial centre $Z$, and that the global subgroup $PG_2'$ is a representation of $(\tilde{G}' \times \tilde{G}')/Z_{diag}$ where $Z_{diag} \cong Z$ is a diagonal subgroup of the centre $Z \times Z$ of $\tilde{G}'^2$. We select two elements $P_1$, $Q_1$ generated by the Lie algebra of the first $\tilde{G}'$ factor ($\tilde{G}'_1$), and two elements $P_2$, $Q_2$ generated by the Lie algebra of the second $\tilde{G}'$ factor ($\tilde{G}'_2$), such that they commute up to a non-trivial element of the centre $Z$ of $\tilde{G}'_i$ \cite{Hooft}:
\bea
P_1 Q_1 & = & z_1 Q_1 P_1 \label{com1}\\
P_2 Q_2 & = & z_2 Q_2 P_2 \label{com2}
\eea
Since we only deal with representations of $(\tilde{G}' \times \tilde{G}')/Z_{diag}$, $z_1$ and $z_2$ are actually elements of the central subgroup $(Z \times Z)/Z_{diag} \cong Z$ . By picking $P_i$ and $Q_i$ in a specific way, we can thus arrange that (\ref{com1}, \ref{com2}) are satisfied with the additional condition 
\be \label{centercond}
z_1 = (z_2)^{-1} \equiv z
\ee
We now construct the commuting elements
\be
P = P_1 P_2, \qquad Q = Q_1 Q_2, \qquad P' = Q_1^n P Q_1^{-n}
\ee    
These elements will serve as candidate holonomies. To make sure that one finds a non-trivial vacuum, one has to calculate the subgroup left unbroken by the holonomies, and verify that its rank is lower than that of the original gauge group. For a more thorough discussion, we refer the reader to \cite{Keur2}, in that paper we discussed results based on subgroups $PG_M'$ with $\widetilde{PG'}_M=SU(2)^M$. We now proceed with results based on subgroups $PG_M'$ with $\widetilde{PG'}_M=SU(N)^M$, for $N > 2$.

\section{Representations of $SU(N > 2)^M$}

Before starting the discussion on explicit realisations of our construction we wish to point out a few subtleties about these groups.

The groups $SU(N>2)$ posses an outer automorphism that corresponds to complex conjugation. In other words, unlike the situation for $SU(2)$, an $SU(N>2)$ irreducible representation (irrep) is in general not equivalent to the complex conjugate of the irrep. The complex conjugated irrep has the same dimension as the original irrep, and in the notation where each irrep is denoted by its dimension, say $\mathbf{n}$, the complex conjugate irrep is denoted by the dimension with a bar on top $\mathbf{\bar{n}}$. It is a matter of convention which irrep is labelled as $\mathbf{n}$ and which as $\mathbf{\bar{n}}$. In the root diagram of the corresponding algebra, this corresponds to an ambiguity in labelling the roots. We will be interested in product groups, where one has this ambiguity in every factor. The choice we will make to label the representations is related to the following issue.

A diagonal subgroup of a group $PG_N'$ can be constructed as follows: construct a Lie algebra ${\cal L}$ for $\tilde{G}'$, consisting of elements $T^a$. Then the Lie algebra for $PG_N'$ has the structure ${\cal L}_1 \oplus {\cal L}_2 \oplus \cdots \oplus {\cal L}_N$ with each of the ${\cal L}_i \cong {\cal L}$. Hence we can write $T^a_i$, for the generator from ${\cal L}_i$ that corresponds to $T^a$ under an isomorphism mapping ${\cal L}$ to ${\cal L}_i$. The diagonal subgroup of $PG_N'$ is then constructed by taking as generators $T^a_1 + T^a_2 + \cdots + T^a_N$. This construction is not unique, there are many isomorphisms from ${\cal L}_i$ to ${\cal L}$, and these will give different diagonal subgroups. All these diagonal subgroups have the same Lie algebra, but they can have a different global structure. As an example consider the $SU(3) \times SU(3)$ irrep $\mathbf{(3, \bar{3})}$, which in a different labelling would be denoted as $\mathbf{(3,3)}$. Constructing the diagonal subgroup of $\mathbf{(3, \bar{3})}$ as outlined in the above, will lead to a tensor representation which decomposes into an $\mathbf{8}$ and a $\mathbf{1}$. The diagonal subgroup of $\mathbf{(3,3)}$ however is a tensor representation which decomposes in a $\mathbf{6}$ and a $\mathbf{\bar{3}}$. We will prefer the labelling $\mathbf{(3, \bar{3})}$ since then the diagonal subgroup (which is actually $SU(3)/\mathbf{Z}_3$) will have a trivial centre.

We construct our holonomies in subgroups that are not simply connected, regardless of the representation of $G$ we are working in. We need to verify case by case that the holonomies that we construct commute in a simply connected representation, implying that they commute in any representation of the gauge group $G$.

For each algebra, there exists a set of ``fundamental weights'' $\Lambda_i$, defined by 
\be \label{fundweight}
 2 \frac{ \langle \Lambda_i , \alpha_j \rangle}{ \langle \alpha_j, \alpha_j \rangle} = \delta_{ij}
\ee
where the $\alpha_j$ are the simple roots of the algebra. $\Lambda_i$ can be written as a linear combination of simple roots with rational coefficients. Each weight is of the form
\be \label{weight}
\lambda = \sum_i (n_i \Lambda_i + m_i \alpha_i) \quad \textrm{with} \ n_i, m_i \in \mathbf{Z}
\ee

When constructing a subalgebra ${\cal L}_{G'}$ of an algebra ${\cal L}_G$, one can arrange that the CSA of ${\cal L}_{G'}$ is contained in the CSA of ${\cal L}_G$. Call the generators of ${\cal L}_{G'}$, $H_{\beta'}$ and $E_{\beta'}$. Then we have $H_{\beta'} = h_{\beta}$ with $h_{\beta} \in {\cal L}_G$. To find the weights of the subgroup, we first find the $h_{\beta}$ associated to simple roots of $G'$. These can be calculated using:
\be
h_{\beta} = H_{\beta'} = [ E_{\beta'}, E_{-\beta'} ]/2
\ee
The weights $\lambda'$ of $G'$ can now be found from the weights $\lambda$ of $G$ with
\be \label{subweight}
\langle \lambda' , \beta' \rangle = \langle \lambda, \beta \rangle
\ee
One way to find out what the congruence class of the representation of $G'$ is, is by expressing the weights $\lambda'$ of the representation of $G'$ in the simple roots $\alpha'_j$ of $G'$. We write
\bd
\lambda' = \sum_k q_k \alpha'_k
\ed
The $q_k$ are easily calculated using (\ref{fundweight}) with $\Lambda'_k$ a fundamental weight of $G'$
\be \label{qk}
2 \langle \Lambda'_k , \lambda' \rangle = q_k \langle \alpha'_k, \alpha'_k \rangle
\ee 
In all cases we will be considering $G'= SU(N)$, and hence $\langle \alpha'_k, \alpha'_k \rangle$ is a $k$-independent number. Since all weights can be obtained from one weight by adding (or subtracting) simple roots, the fractional part of the $q_k$ is sufficient to determine the weight lattice, and thereby the congruence class of the representation.

We will now continue with the explicit constructions. Our conventions are outlined in the appendices of \cite{Keur2}. For the decompositions of groups into subgroups, use was made of \cite{McKay}.

\section{Non-trivial vacua based on twist in $SU(3)$} 

We will start by developing the relevant tools for this group. After that we will give an overview of groups in which our construction can be realised. These are all the exceptional groups except $G_2$. $F_4$ will be discussed in quite some detail.

Like in \cite{Keur2} where $\tilde{G}' = SU(2)$ was discussed, our construction can always be carried out in a subgroup $SU(3)^2$, but we will often take $SU(3)^M$ with $M > 2$, as this allows us to choose it to be a regular subgroup. This will give important simplifications in the calculations.

\subsection{Twist in $SU(3)$}

We will take the canonical form (in the conventions of the appendices of \cite{Keur2}, with the modification that we will use capital $E$ and $H$ to be able to distinguish the subgroup generated by these from the original group, whose generators we keep on denoting by $h_{\alpha}$ and $e_{\alpha}$):
\be
\begin{array}{lll}
\lbrack H_{\alpha_1}, E_{\alpha_1} \rbrack = \frac{1}{3} E_{\alpha_1} & \lbrack H_{\alpha_1}, E_{\alpha_2} \rbrack = -\frac{1}{6} E_{\alpha_1} & \lbrack H_{\alpha_1}, E_{\alpha_1+ \alpha_2} \rbrack = \frac{1}{6} E_{\alpha_1+\alpha_2} \\
\lbrack H_{\alpha_2}, E_{\alpha_1} \rbrack = -\frac{1}{6} E_{\alpha_1} & \lbrack H_{\alpha_2}, E_{\alpha_2} \rbrack = \frac{1}{3} E_{\alpha_2} & \lbrack H_{\alpha_2}, E_{\alpha_1+ \alpha_2} \rbrack = \frac{1}{6} E_{\alpha_1+\alpha_2} \\
\lbrack E_{\alpha_1}, E_{-\alpha_1} \rbrack = 2 H_{\alpha_1} & \lbrack E_{\alpha_2}, E_{-\alpha_2} \rbrack = 2 H_{\alpha_2} & \lbrack E_{\alpha_1}, E_{\alpha_2} \rbrack = \frac{1}{\sqrt{3}} E_{\alpha_1+ \alpha_2} \\
\multicolumn{3}{c}{\lbrack E_{\alpha_1+\alpha_2}, E_{-\alpha_1- \alpha_2} \rbrack = 2 H_{\alpha_1+\alpha_2} = 2 H_{\alpha_1}+ 2 H_{\alpha_2} }
\end{array}
\ee
All other relations can be found by conjugation, and the Jacobi identity.

In $SU(3)$ we look for two matrices satisfying
\be
pq = \exp(\frac{2 \pi i}{3})qp
\ee
$p$ and $q$ will commute when lifted to $SU(3) / \mathbf{Z}_3$. We take:
\be
p  =  \left( \begin{array}{ccc} \exp(\frac{2 \pi i}{3}) & 0 & 0  \\ 0 & 1 & 0 \\
0 & 0 & \exp(-\frac{2 \pi i}{3}) \end{array} \right),  \qquad
q  = \left( \begin{array}{ccc} 0 & 1 & 0  \\ 0 & 0 & 1 \\ 1 & 0 & 0 \end{array} \right) \ee
In terms of generators this is
\bea
p & = & \exp \left( 4 \pi i H_{\alpha_1 + \alpha_2} \right) \label{pqsu3} \\ q  & = & \exp \left( \frac{4 \pi i}{3} \frac{1}{2i}(E_{\alpha_1} + E_{\alpha_2} + E_{-\alpha_1 - \alpha_2} - E_{-\alpha_1} - E_{-\alpha_2} - E_{\alpha_1 + \alpha_2}) \right) \non
\eea
The commutation relations of $p$ and $q$ with the group generators are most easily calculated in a specific representation. One finds
\be
\begin{array}{rclcrcl}
p H_{\alpha} & = & H_{\alpha} p, & & q H_{\alpha} & = & H_{R \alpha} q, \\
p E_{\alpha} & = & \exp (4 \pi i \langle \alpha, \alpha_1 + \alpha_2  \rangle) E_{\alpha} p, & & q E_{\alpha} & = & E_{R \alpha} q
\end{array}
\ee

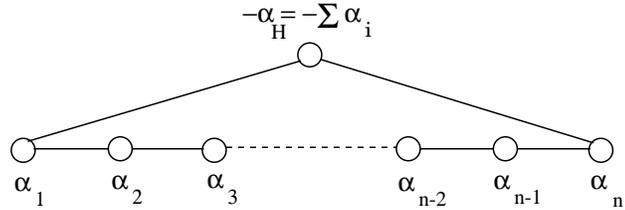
\EPSFIGURE{edynka.eps}{The extended Dynkin diagram for $SU(N+1)$}

The action of the rotation $R$ (This is a genuine rotation over $2 \pi / 3$ in the root diagram) is fully determined by its action on the simple roots.
$R$ is an element of the Weyl group: It is the composition of the Weyl reflection generated by $\alpha_2$, followed by the reflection generated by $\alpha_1$. The effect of $R$ is
\be \label{rot3}
R : \qquad \alpha_1 \rightarrow \alpha_2 \rightarrow -(\alpha_1 + \alpha_2) \rightarrow \alpha_1
\ee

A nice mnemonic for the action of $R$ is provided by the ``extended Dynkin diagram'' \cite{Dynkin} of $SU(3)$. The extended Dynkin diagram of an algebra consists of its Dynkin diagram, extended with one more root, $-\alpha_H$, with $\alpha_H$ the highest root of the algebra. The dot representing the highest root is then connected to the diagram via the standard rules. For $su(3)$, $-\alpha_H = -(\alpha_1 + \alpha_2)$ and the extended diagram consists of 3 dots, connected to form a cycle. The action of $R$ is nothing but a rotation of this cycle by one step. For twist in the higher unitary groups we will see the same feature: the action of the analogon of the element $q$ can be represented by rotating the (cyclic) extended Dynkin diagram of the unitary group by one step.

\subsection{Calculation of the unbroken subgroup}

Again we wish to compute the combinations $pT^ap^{-1}$ and $qT^aq^{-1}$, with $T^a$ the generators of $G$. We will  generalise our methods for the $SU(2)$-case to $SU(3)$, but this is complicated by a crucial difference between $SU(2)$ and $SU(3)$: For $SU(2)$ all weights in an irreducible representation have multiplicity $1$, whereas this is not true for $SU(3)$-irreps. However, we will need only three irreps of $SU(3)$: the $\mathbf{1}$ and $\mathbf{10}$, which do contain only simple weights, and the $\mathbf{8}$, which is equivalent to the adjoint. Note that these are all irreps of $SU(3)/\mathbf{Z}_3$.

We proceed as in the $SU(2)$-case. $pT^ap^{-1}$ is easily calculated, with $p = \exp(i h_{\zeta})$ we have
\be \label{hcomm}
p h_{\alpha} p^{-1} = h_{\alpha} \qquad p e_{\beta} p^{-1} = \exp( i \langle \zeta, \beta \rangle) e_{\beta}
\ee
To calculate $qT^aq^{-1}$ we start by decomposing the representation of the group $G$ into irreducible representations of $SU(3)/\mathbf{Z_3}$. First we construct the normalised eigenvectors $\psi_{\lambda}$ of the CSA: $H_{\alpha} \psi_{\lambda} = \langle \alpha, \lambda \rangle \psi_{\lambda}$. It then follows that
\be 
H_{R \alpha} q \psi_{\lambda} =  q H_{\alpha} \psi_{\lambda} = \langle \alpha , \lambda \rangle q \psi_{\lambda} = \langle R \alpha, R \lambda \rangle q \psi_{\lambda}
\ee
This means that $q \psi_{\lambda}$ is an eigenvector of the CSA with weight $R \lambda$. If all weights are simple, as in the $\mathbf{1}$ and $\mathbf{10}$, then everything is easily solvable, as in the $SU(2)$ case: $q \psi_{\lambda} = \phi_{\lambda} \psi_{R \lambda}$ with $\phi_{\lambda}$ a phase. From $q^3 = 1$, we find that $\phi_{\lambda} \phi_{R \lambda} \phi_{R^2 \lambda} = 1$. With the ladder operators one shows that $\phi_{\lambda}$ is actually independent of $\lambda$ (in a suitable phase convention). The following three combinations are eigenvectors:
\bd
\begin{array}{rcl}
\psi_{\lambda} + \psi_{R \lambda} + \psi_{R^2 \lambda} & \rightarrow & \textrm{eigenvalue } \phi \\
\psi_{\lambda} + \exp(\frac{4 \pi i}{3}) \psi_{R \lambda} + \exp(\frac{2 \pi i}{3}) \psi_{R^2 \lambda} & \rightarrow & \textrm{eigenvalue } \exp(\frac{2 \pi i}{3}) \phi \\
\psi_{\lambda} + \exp(\frac{2 \pi i}{3}) \psi_{R \lambda} + \exp(\frac{4 \pi i}{3}) \psi_{R^2 \lambda} & \rightarrow & \textrm{eigenvalue } \exp(\frac{4 \pi i}{3}) \phi 
\end{array}
\ed
Note that formally we have to turn to the complexification of the real Lie-algebra for this to make sense. For $\lambda = 0$ the last two of these combinations are zero. We find that each triple of eigenvectors contributes $\phi^3=1$ to the determinant, and since all weights, apart from the zero weight, occur in triples we find that the determinant is equal to $\phi^{3n+1} = \phi = 1$.

When not all weights have multiplicity 1, the above discussion for the non-zero weights applies anyway. It is the zero weight that causes the problems. The eigenvectors with weight zero form a subspace that will be mapped into itself, our previous methods fail, and we know no easy way to determine the action of $q$ on any vector of this subspace. For the $\mathbf{8}$, the solution is nevertheless easy to find by realising that this is the adjoint. The zero weights of the adjoint representation are the generators of the CSA, the $H_{\alpha}$, on which we already know the action of $q$, and the eigenvectors of $q$ are
\bd
\begin{array}{rcl}
H_{\alpha} + H_{R \alpha} + H_{R^2 \alpha} & = & 0 \\
H_{\alpha} + \exp(\frac{4 \pi i}{3}) H_{R \alpha} + \exp(\frac{2 \pi i}{3}) H_{R^2 \alpha} & \rightarrow & \textrm{eigenvalue } \exp(\frac{2 \pi i}{3}) \\
H_{\alpha} + \exp(\frac{2 \pi i}{3}) H_{R \alpha} + \exp(\frac{4 \pi i}{3}) H_{R^2 \alpha} & \rightarrow & \textrm{eigenvalue } \exp(\frac{4 \pi i}{3}) 
\end{array}
\ed
We can now construct the action of $q$ on any representation of $G$ that splits into singlets, octets and decuplets. As a final remark we note that these considerations imply that
\bd
p T^a p^{-1} = \exp (\frac{2 \pi i n}{3}) \ T^a, \qquad qT^aq^{-1} = \exp (\frac{2 \pi i m}{3}) \ T^a
\ed

For the $su(3)$-subalgebra, we define the CSA generators $H_{\alpha'_1}$ and $H_{\alpha'_2}$ associated to the simple roots of $su(3)$, following the conventions of the appendix of \cite{Keur2}. Set $H_{\alpha'_1}= h_{\beta_1}$, $H_{\alpha'_2} = h_{\beta_2}$ for $h_{\beta_i} \in {\cal L}_G$. We need the combinations $2 \Lambda'_i / \langle \alpha'_k, \alpha'_k \rangle$ (see eq. (\ref{qk})), where $\Lambda'_i$ are the fundamental weights of $su(3)$ and $\alpha'_k$ are simple roots of $su(3)$. In our conventions $\langle \alpha'_k, \alpha'_k \rangle = 1/3$, for $\alpha'_k$ a root of $su(3)$. The combinations we need are thus 
\bd
 6 \Lambda'_1 = 4 \beta_1 + 2 \beta_2 \qquad 6 \Lambda'_2 = 2 \beta_1 + 4 \beta_2.
\ed 
We write the weights of the representation of the group $G$ as (see \ref{weight})
\bd
\lambda = \sum_i (n_i \Lambda_i + m_i \alpha_i) \quad \textrm{with} \ n_i, m_i \in \mathbf{Z}
\ed
with $\Lambda_i$ the fundamental weights of $G$ and $\alpha_i$ its simple roots. We will demand that the representation of $G$ branches into representations of $SU(3)/\mathbf{Z}_3$, a condition that can be translated to
\be \label{su3cond}
\langle \alpha_i , 4 \beta_1 + 2 \beta_2 \rangle, \quad
\langle \alpha_i , 2 \beta_1 + 4 \beta_2 \rangle, \quad
\langle \Lambda_i , 4 \beta_1 + 2 \beta_2 \rangle, \quad
\langle \Lambda_i , 2 \beta_1 + 4 \beta_2 \rangle  \quad \in \mathbf{Z} 
\ee

\subsection{Realisation of the $SU(3)$-based construction}

\subsubsection{$F_4$} \label{f4nontriv}

$F_4$ possesses a $PG'_2$-subgroup with $\widetilde{PG'}_2= SU(3)^2$. The roots of the subalgebra are contained in the $f_4$ root lattice. An $su(3)$-sublattice is completely determined by making a choice for its simple roots. We take for the first $su(3)$ factor the two roots $(1,-1,0,0)/\sqrt{18}$ and $(0,1,-1,0)/\sqrt{18}$, which are of equal length and have an angle in-between of $2 \pi /3$ and can therefore be used as simple roots for $su(3)$. For the second $su(3)$ factor we take as simple roots $(0,0,0,1)/\sqrt{18}$ and $(1,1,1,-1)/2\sqrt{18}$. These two are orthogonal to the root vectors of the first $su(3)$. To construct the $su(3)$ algebra's, we need to include appropriate normalisation factors. The first $SU(3)$-factor is generated by:
\bea
H_{\alpha_1}^1 = 3 h_{(1,-1,0,0)/\sqrt{18}} & H_{\alpha_2}^1 =  3 h_{(0,1,-1,0)/\sqrt{18}} & \\
E_{\alpha_1}^1 = \sqrt{3} e_{(1,-1,0,0)/\sqrt{18}} &  E_{\alpha_2}^1 =  \sqrt{3} e_{(0,1,-1,0)/\sqrt{18}} & E_{\alpha_1 + \alpha_2}^1 =  \sqrt{3} e_{(1,0,-1,0)/\sqrt{18}} \non
\eea
The second $SU(3)$ factor is generated by
\bea
H_{\alpha_1}^1 = 6 h_{(0,0,0,1)/\sqrt{18}} & H_{\alpha_2}^1 =  6 h_{(1,1,1,-1)/2\sqrt{18}} & \\
E_{\alpha_1}^1 = \sqrt{6} e_{(0,0,0,1)/\sqrt{18}} &  E_{\alpha_2}^1 =  \sqrt{6} e_{(1,1,1,-1)/2\sqrt{18}} & E_{\alpha_1 + \alpha_2}^1 =  \sqrt{6} e_{(1,1,1,1)/2\sqrt{18}} \non
\eea

The diagonal subalgebra $D$ is then easily constructed:
\bea
H_{\alpha_1}^D = & H_{\alpha_1}^1 + H_{\alpha_1}^2 & = \frac{1}{2} \sqrt{2} h_{(1,-1,0,2)} \non \\
H_{\alpha_2}^D = & H_{\alpha_2}^1 + H_{\alpha_2}^2 & = \frac{1}{2} \sqrt{2} h_{(1,2,0,-1)} \non \\
E_{\alpha_1}^D = & E_{\alpha_1}^1 + E_{\alpha_1}^2 & = \sqrt{3} e_{(1,-1,0,0)/\sqrt{18}} + \sqrt{6} e_{(0,0,0,1)/\sqrt{18}}   \\
E_{\alpha_2}^D = & E_{\alpha_2}^1 + E_{\alpha_2}^2 & =  \sqrt{3} e_{(0,1,-1,0)/\sqrt{18}} + \sqrt{6} e_{(1,1,1,-1)/2\sqrt{18}}\non \\
E_{\alpha_1 + \alpha_2}^D = & E_{\alpha_1+ \alpha_2}^1 + E_{\alpha_1 + \alpha_2}^2 & =  \sqrt{3} e_{(1,0,-1,0)/\sqrt{18}} + \sqrt{6} e_{(1,1,1,1)/2\sqrt{18}}\non 
\eea
This algebra generates an $SU(3)/\mathbf{Z}_3$-group, as can be seen by checking the conditions (\ref{su3cond}). We have $4 \alpha_1 + 2 \alpha_2 = \sqrt{18}(1,0,0,1)$ and $2 \alpha_1 + 4 \alpha_2 = \sqrt{18}(1,1,0,0)$, and the inner product of these vectors with the simple roots of $F_4$ all give integers. $F_4$ has no non-integer fundamental weights. The decomposition of $F_4$ into the diagonal $SU(3)/\mathbf{Z}_3$ is as follows:
\bea
F_4 \quad \rightarrow & SU(3) \times SU(3) & \rightarrow \quad SU(3) / \mathbf{Z}_3 \non \\
\mathbf{26} \quad \rightarrow & \mathbf{(3,\bar{3}) \oplus (\bar{3},3) \oplus (8,1)} & \rightarrow \quad 3\mathbf{(8)} \oplus 2\mathbf{(1)}\\
\mathbf{52} \quad \rightarrow & \mathbf{(8,1) \oplus (6,3) \oplus (\bar{6},\bar{3}) \oplus \mathbf(1,8)} & \rightarrow \quad \mathbf{(10)} \oplus \mathbf{(\overline{10})} \oplus 4 \mathbf{(8)} \non 
\eea

Now look for two matrices that commute, but cannot be written as exponentials of generators of $D$ that commute. From (\ref{pqsu3})
\bea
P & = & \exp \left( 4 \pi i H_{\alpha_1 + \alpha_2}^D \right) \\ Q  & = & \exp \left( \frac{4 \pi i}{3} \frac{1}{2i}(E_{\alpha_1}^D + E_{\alpha_2}^D + E_{-\alpha_1 - \alpha_2}^D - E_{-\alpha_1}^D - E_{-\alpha_2}^D - E_{\alpha_1 + \alpha_2}^D) \right) \non
\eea
Splitting the generators of the diagonal group as $H^D_{\alpha} = H^1_{\alpha} + H^2_{\alpha}$ and $E^D_{\alpha} = E^1_{\alpha} + E^2_{\alpha}$ reveals the product structure of $P$ and $Q$, which decompose accordingly as $P = P_1 P_2$, and $Q = Q_1 Q_2$. That $P$ and $Q$ commute follows from the fact that they take the role of the elements $p$ and $q$ constructed previously, and that they are elements of an $SU(3)/\mathbf{Z}_3$-subgroup. Now we wish to construct $P'$.  Our methods for $SU(2)$ \cite{Keur2} have to be generalised in such a way, that it produces a subgroup $D'$ that is isomorphic to $D$, such that $Q$ is an element of both $D$ and $D'$. The appropriate generalisation has the form of a rotation of one of the two factors
\be \label{twist1}
H_{\alpha}^1 \rightarrow H_{R \alpha}^1 \qquad E_{\alpha}^1 \rightarrow E_{R \alpha}^1
\ee
where $R$ is the rotation (\ref{rot3}). The second $SU(3)$ will be left as it is. We now construct the diagonal group $D'$:
\bea
H_{\alpha_1}^{D'} = & H_{\alpha_2}^1 + H_{\alpha_1}^2 & = \frac{1}{2} \sqrt{2} h_{(0,1,-1,2)} \non \\
H_{\alpha_2}^{D'} = & H_{-\alpha_1 - \alpha_2}^1 + H_{\alpha_2}^2 & = \frac{1}{2} \sqrt{2} h_{(0,1,2,-1)} \non \\
E_{\alpha_1}^{D'} = & E_{\alpha_2}^1 + E_{\alpha_1}^2 & = \sqrt{3} e_{(0,1,-1,0)/\sqrt{18}} + \sqrt{6} e_{(0,0,0,1)/\sqrt{18}}   \\
E_{\alpha_2}^{D'} = & E_{-\alpha_1 -\alpha_2}^1 + E_{\alpha_2}^2 & =  \sqrt{3} e_{(-1,0,1,0)/\sqrt{18}} + \sqrt{6} e_{(1,1,1,-1)/2\sqrt{18}}\non \\
E_{\alpha_1 + \alpha_2}^{D'} = & E_{-\alpha_1}^1 + E_{\alpha_1 + \alpha_2}^2 & =  \sqrt{3} e_{(-1,1,0,0)/\sqrt{18}} + \sqrt{6} e_{(1,1,1,1)/2\sqrt{18}}\non 
\eea
We set $P'$ to be
\be
P' = \exp( 4 \pi i H_{\alpha_1 + \alpha_2}^{D'}) 
\ee
$P$, $P'$ and $Q$ commute by construction. Since $F_4$ is simply connected, the gauge connection implied by the holonomies $\Omega_1= P$, $\Omega_2=P'$ and $\Omega_3= Q$ is flat and non-trivial.

Now we calculate the unbroken subgroup: first find the generators that commute with $P$, of these check how many commute with $P'$, and finally compute the commutator with $Q$ of the generators that commute with both $P$ and $P'$. These computations can be done by using our previous results and substituting $P$ for $p$ and $Q$ for $q$ in the first step, and in a second step $P'$ for $p$ and $Q$ for $q$ . The only $e_{\pm \beta}$ that commute with $P$ have 
\bd
\begin{array}{c}
\beta \in \{ \frac{(0,0,1,0)}{\sqrt{18}}, \frac{(1,1,0,0)}{\sqrt{18}}, \frac{(1,0,0,1)}{\sqrt{18}}, \frac{(0,1,0,-1)}{\sqrt{18}}, \frac{(1,-1,1,-1)}{2 \sqrt{18}}, \frac{(1,-1,-1,-1)}{2 \sqrt{18}} \} 
\end{array}
\ed
However, none of these commute with $P'$. Hence the CSA-generators are the only ones that commute with both $P$ and $P'$. The effect of $Q$ on the CSA can be computed by studying the branching of $F_4$ into $SU(3)/\mathbf{Z}_3$. Careful examination shows that all CSA-generators have weight zero for the $\mathbf{8}$'s in the decomposition. There is no combination of zero weights for the $\mathbf{8}$ that is invariant under the adjoint action of $Q$ and hence no generator commutes with $P$, $P'$ and $Q$. The unbroken subgroup is therefore discrete.

We therefore have found a \emph{new} flat connection: Neither the trivial flat connection nor the flat connection constructed in \cite{Keur2} can be deformed to a flat connection that has a discrete unbroken subgroup. Note that instead of the rotation (\ref{twist1}) we could also have rotated by
\be \label{twist2}
H_{\alpha}^1 \rightarrow H_{R^2 \alpha}^1 \qquad E_{\alpha}^1 \rightarrow E_{R^2 \alpha}^1
\ee
Repeating the steps for the construction of $D'$, this gives another diagonal subgroup, that we will call $D''$. It is also possible to construct an element $P'' =  \exp( 4 \pi i H_{\alpha_1 + \alpha_2}^{D''})$. $P''$ commutes with $P$, $P'$, and $Q$, and can be used in the construction of non-trivial flat connections. The flat connection corresponding to $\Omega_1 = P$, $\Omega_2 = P'$ and $\Omega_3 = Q$ is not equivalent to the one corresponding to $\Omega_1 = P$, $\Omega_2 = P''$ and $\Omega_3 = Q$, as will be proven in a subsequent publication. These two flat connections both have discrete unbroken subgroups. 

We remind the reader that the non-trivial flat connection in $SO(7)$ in \cite{Witnew} was characterised by simultaneously diagonalising the three holonomies in the 7 dimensional vector representation. The 7 triples $((\Omega_1)_{ii}, (\Omega_2)_{ii}, (\Omega_3)_{ii})$ are the triples $( \pm 1, \pm 1, \pm 1)$ with at least one $-1$. In Witten's original construction, the triples $( \pm 1, \pm 1, \pm 1)$ correspond to the positions of D-branes at orientifold fixed points. 

There is a remarkable resemblance between this result and the $SU(3)$-based non-trivial flat connection in $F_4$. We can diagonalise the holonomies constructed here. On the diagonal we have the eigenvalues $\exp (2 \pi i n /3)$. Take now the triples 
\be \label{f4triples}
( \exp(\frac{2 \pi i n_1}{3}),\exp( \frac{2 \pi i n_2}{3}),\exp(\frac{2 \pi i n_3}{3})) \qquad n_i \in \mathbf{Z} 
\ee
and exclude the triple $(1,1,1)$. There are $3^3-1 =26$ distinct triples, and this is precisely the dimension of the fundamental irrep of $F_4$. Indeed, constructing the elements $P$, $P'$ and $Q$ in the $\mathbf{26}$ of $F_4$ and diagonalising, we find that the triples of diagonal elements $(P_{ii}, P'_{ii}, Q_{ii})$ are precisely the 26 triples mentioned above. It is an intriguing question whether this too can be related to an M-theory construction.  

\subsubsection{$E_6$}

In $E_6$ there is a $SU(3)^3$-subgroup that is suitable for our purposes. The $su(3)$-factors have root vectors
\be \label{e6su3roots}
\begin{array}{rll}
su(3)_1: & \alpha_1^1 = (0,0,0,0,1,1)/\sqrt{24}, & \alpha_2^1 = (\sqrt{3},-1,-1,-1,-1,-1)/2 \sqrt{24}; \\
su(3)_2: & \alpha_1^2 = (0,0,1,-1,0,0)/\sqrt{24}, & \alpha_2^2 = (0,1,-1,0,0,0)/\sqrt{24}; \\
su(3)_3: & \alpha_1^3 = (0,0,0,0,1,-1)/\sqrt{24}, & \alpha_2^3 = (-\sqrt{3},-1,-1,-1,-1,1)/2 \sqrt{24}. 
\end{array}
\ee
The relevant geometrical properties are easily verified. To construct the $su(3)$ algebra's, we need to include appropriate normalisation factors. A full exposition would be very space consuming, so we limit ourselves to a few points.

The diagonal subalgebra $D$ is generated by:
\bea
H_{\alpha_1}^D = & H_{\alpha_1}^1 + H_{\alpha_1}^2 + H_{\alpha_1}^3 & = \frac{1}{3} \sqrt{6} h_{(0,0,1,-1,2,0)} \non \\
H_{\alpha_2}^D = & H_{\alpha_2}^1 + H_{\alpha_2}^2 + H_{\alpha_1}^3 & = \frac{1}{3} \sqrt{6} h_{(0,0,-2,-1,-1,0)} \non \\
E_{\alpha_1}^D = & E_{\alpha_1}^1 + E_{\alpha_1}^2 + E_{\alpha_1}^3 & = 2 (e_{\alpha_1^1} + e_{\alpha_1^2} + e_{\alpha_1^3})  \\
E_{\alpha_2}^D = & E_{\alpha_2}^1 + E_{\alpha_2}^2 + E_{\alpha_2}^3 & = 2 (e_{\alpha_2^1} + e_{\alpha_2^2} + e_{\alpha_2^3})  \non \\
E_{\alpha_1 + \alpha_2}^D = & E_{\alpha_1 + \alpha_2}^1 + E_{\alpha_1 + \alpha_2}^2 + E_{\alpha_1+\alpha_2}^3 & = 2 (e_{\alpha_1^1+ \alpha_2^1} + e_{\alpha_1^2 + \alpha_2^2} + e_{\alpha_1^3 + \alpha_2^3})  \non \\
\eea
This algebra generates an $SU(3)/\mathbf{Z}_3$-group, as can be seen by checking the conditions (\ref{su3cond}). This subgroup corresponds to the decompositions
\bea
E_6 \quad \rightarrow & SU(3) \times SU(3) \times SU(3) & \rightarrow \quad SU(3) / \mathbf{Z}_3 \non \\
\mathbf{27} \quad \rightarrow & \mathbf{(3,\bar{3},1) \oplus (\bar{3},1,3) \oplus (1,3, \bar{3})} & \rightarrow \quad 3\mathbf{(8)} \oplus 3\mathbf{(1)}\\
\mathbf{78} \quad \rightarrow & \mathbf{(8,1,1) \oplus (1,8,1) \oplus (1,1,8)} & \non \\
&  \mathbf{\oplus (3,3,3) \oplus (\bar{3},\bar{3}, \bar{3})} & \rightarrow \quad \mathbf{(10)} \oplus \mathbf{(\overline{10})} \oplus 7 \mathbf{(8)} \oplus 2 \mathbf{(1)} \non 
\eea
The diagonal subalgebra's $D'$ and $D''$ are obtained by rotating one of the three $SU(3)$ factors by (\ref{twist1}) resp. (\ref{twist2}). The holonomies are then  $\Omega_1 = P$, $\Omega_2 = P'$ and $\Omega_3 = Q$ with
\be
\begin{array}{ccc}
P = \exp (4 \pi i H_{\alpha_1 + \alpha_2}^D) & P' = \exp( 4 \pi i H_{\alpha_1 + \alpha_2}^{D'}) \\
\multicolumn{2}{c}{Q = \exp (\frac{4 \pi i}{3} \frac{1}{2i}(E_{\alpha_1}^D + E_{\alpha_2}^D + E_{-\alpha_1 - \alpha_2}^D - E_{-\alpha_1}^D - E_{-\alpha_2}^D - E_{\alpha_1 + \alpha_2}^D))}
\end{array}
\ee
Again the subgroup $D''$ gives rise to an element $P''$, and the exchange $P' \leftrightarrow P''$ produces a non-equivalent flat connection with the unbroken discrete subgroup.

These $E_6$-flat connections are essentially the same as the ones for $F_4$, as can be understood from the decomposition.
\bea
E_6 & \rightarrow & F_4 \non \\
\mathbf{27} & \rightarrow & \mathbf{26 \oplus 1} \\
\mathbf{78} & \rightarrow & \mathbf{52 \oplus 26} \non 
\eea
Like the embedding of $G_2$ in $SO(7)$ of \cite{Keur2}, it is not hard to make this explicit. Take the roots of $E_6$ (in the conventions of \cite{Keur2}), set the first and last component of each vector to zero, and renormalise the vectors by multiplying with $2 / \sqrt{3}$. This projection gives the roots of $F_4$. Applying the same projection to the roots (\ref{e6su3roots}), one finds that the $SU(3)^3$ subgroup of $E_6$ projects onto an $SU(3)^2$ subgroup of $F_4$.

We furthermore note that, concerning the remark in the previous paragraph on the triples (\ref{f4triples}), that in the fundamental $\mathbf{27}$ irrep of $E_6$, the holonomies can be constructed from these triples with the triple $(1,1,1)$ \emph{included}.

\subsubsection{$E_7$}

Because the $E_6$-lattice is a sub-lattice of the $E_7$-lattice, it is not hard to embed the previous $E_6$ result into $E_7$. Working out all details, one finds two inequivalent non-trivial flat connections with an unbroken $SU(2)$, precisely as one would naively expect from the decomposition of $E_7$ in $F_4 \times SU(2)$
\bea
E_7 & \rightarrow & F_4 \times SU(2) \non \\
\mathbf{56} & \rightarrow & \mathbf{(26,2) \oplus (1,4)} \\
\mathbf{133} & \rightarrow & \mathbf{(52,1) \oplus (26,3) \oplus (1,3)} \non 
\eea

\subsubsection{$E_8$}

As the $E_6$-lattice is also a sublattice of the lattice of $E_8$, our construction is easily extended to $E_8$. One finds two inequivalent non-trivial flat connections with unbroken subgroup $G_2$, in accordance to  the decomposition
\bea
E_8 & \rightarrow & G_2 \times F_4 \non \\
\mathbf{248} & \rightarrow & \mathbf{(14,1) \oplus (7,26) \oplus (1,52)} \label{e8g2f4}
\eea

\section{Non-trivial vacua based on twist in $SU(4)$} 

It is also possible to use $SU(4)$ for a construction of non-trivial flat connections. However, we need a subtle modification with respect to the $SU(2)$ and $SU(3)$ constructions.  After developing the relevant tools for $SU(4)$, we will construct the explicit realisation of this construction in the groups $E_7$ and $E_8$. 

\subsection{Twist in $SU(4)$}

We will take the canonical form (in the conventions of the appendices of \cite{Keur2}), where again we will use $H_{\alpha}$ and $E_{\alpha}$ for the generators of the subgroup, and $h_{\alpha}$ and $e_{\alpha}$ for the generators of the original group.

In $SU(4)$ we look for two matrices satisfying
\be
pq = \exp(\frac{2 \pi i}{4})qp
\ee
We take:
\be
p = \left( \begin{array}{cccc} \exp(\frac{3 \pi i}{4}) & 0 & 0 & 0  \\ 0 & \exp(\frac{\pi i}{4}) & 0 & 0 \\ 0 & 0 & \exp(-\frac{\pi i}{4}) & 0 \\
0 & 0 & 0 & \exp(-\frac{3 \pi i}{4}) \end{array} \right) \qquad 
q =  \left( \begin{array}{cccc} 0 & 1 & 0 & 0 \\ 0 & 0 & 1  & 0 \\ 0 & 0 & 0 & 1 \\ 1 & 0 & 0 & 0 \end{array} \right) \exp (\frac{\pi i}{4})
\ee

In terms of generators this is
\bea
p & = & \exp \left( 2 \pi i  H_{3 \alpha_1 + 4 \alpha_2 + 3 \alpha_3} \right) \\
q & = & \exp \Bigl( \frac{\pi i}{2}( (1-i)(E_{\alpha_1} + E_{\alpha_2} + E_{\alpha_3} + E_{-\alpha_1- \alpha_2 - \alpha_3}) -(E_{\alpha_1+ \alpha_2} + E_{\alpha_2+ \alpha_3}) \\ & & \qquad -(E_{-\alpha_1-\alpha_2} + E_{-\alpha_2 - \alpha_3}) + (1+i)(E_{-\alpha_1} + E_{-\alpha_2} + E_{-\alpha_3} + E_{\alpha_1 + \alpha_2 + \alpha_3})) \Bigr) \non
\eea

The commutation relations of $p$ and $q$ with the group generators are most easily calculated in a specific representation. One finds
\be
\begin{array}{rclcrcl}
p H_{\alpha} & = & H_{\alpha} p, & & q H_{\alpha} & = & H_{R \alpha} q, \\
p E_{\alpha} & = & \exp (2 \pi i \langle \alpha, 3\alpha_1 + 4\alpha_2 + 3\alpha_4  \rangle) E_{\alpha} p, & & q E_{\alpha} & = & E_{R \alpha} q.
\end{array}
\ee
The action of the rotation $R$ (which is in this case a combination of a genuine rotation and a reflection) is fully determined by its action on the simple roots.
\be
R : \qquad \alpha_1 \rightarrow \alpha_2 \rightarrow \alpha_3 \rightarrow -(\alpha_1 + \alpha_2 + \alpha_3) \rightarrow \alpha_1
\ee
Just like in the $SU(3)$-case this is nicely visualised by a rotation of the extended Dynkin diagram of $SU(4)$. $R$ is an element of the Weyl group: It is the composition of the Weyl reflection generated by $\alpha_3$, followed by the reflection generated by $\alpha_2$ and the reflection generated by $\alpha_1$. It is obvious that $R^4 = 1$, but on the vectors that are a multiple of $\alpha_1 + \alpha_3$  $R$ acts as a reflection, and we even have $R^2 (\alpha_1 + \alpha_3) = \alpha_1 + \alpha_3$. 

\subsection{Calculation of the unbroken subgroup}

For the computation of the combinations $pT^ap^{-1}$ and $qT^aq^{-1}$, we will need several representations of $SU(4)$. For the construction in the fundamental $\mathbf{56}$ irrep of $E_7$ we need the $\mathbf{1}$, $\mathbf{15}$ and $\mathbf{6}$ of $SU(4)$. For the construction in the adjoints of $E_7$ and $E_8$ (the $\mathbf{133}$ resp. $ \mathbf{248}$) we furthermore need the $\mathbf{10}$ and $\mathbf{20}$ of $SU(4)$. Note that the $\mathbf{6}$ and $\mathbf{10}$ still contain a non-trivial $\mathbf{Z}_2$ centre, which is a subgroup of the $\mathbf{Z}_4$ centre of their simply connected covering $SU(4)$. To compensate for the non-trivial $\mathbf{Z}_2$, we will use the $SU(2)$ factor in the decomposition $E_7 \rightarrow SU(4) \times SU(4) \times SU(2)$. We will need only the one-, two- and three dimensional representation of $SU(2)$, and relevant facts about these representations can be found in our previous article \cite{Keur2}.

The calculation of $pT^ap^{-1}$ proceeds in the same way as in (\ref{hcomm}). For $qT^aq^{-1}$ we start again by decomposing the representation of the group $G$ into irreducible representations of $SU(4)$. Construct the normalised eigenvectors $\psi_{\lambda}$ of the CSA: $H_{\alpha} \psi_{\lambda} = \langle \alpha, \lambda \rangle \psi_{\lambda}$. It then follows that
\be 
H_{R \alpha} q \psi_{\lambda} =  q H_{\alpha} \psi_{\lambda} = \langle \alpha , \lambda \rangle q \psi_{\lambda} = \langle R \alpha, R \lambda \rangle q \psi_{\lambda}
\ee
This means that $q \psi_{\lambda}$ is an eigenvector of the CSA with weight $R \lambda$. For simple weights, we have $q \psi_{\lambda} = \phi_{\lambda} \psi_{R \lambda}$ with $\phi_{\lambda}$ a phase. Because all representations we consider are representations of $SU(4)/\mathbf{Z}_2$, we have $q^4 = 1$ and we find that $\phi_{\lambda} \phi_{R \lambda} \phi_{R^2 \lambda} \phi_{R^3 \lambda} = 1$. With the ladder operators one shows that $\phi_{\lambda}$ is actually independent of $\lambda$ (in a suitable phase convention). For $\lambda \neq 0$ the following four combinations are eigenvectors:
\bd
\begin{array}{rcl}
\psi_{\lambda} + \psi_{R \lambda} + \psi_{R^2 \lambda} + \psi_{R^3 \lambda} & \rightarrow & \textrm{eigenvalue } \phi \\
\psi_{\lambda} - i \psi_{R \lambda} - \psi_{R^2 \lambda} + i \psi_{R^3 \lambda} & \rightarrow & \textrm{eigenvalue } i \phi \\
\psi_{\lambda} - \psi_{R \lambda} + \psi_{R^2 \lambda} - \psi_{R^3 \lambda} & \rightarrow & \textrm{eigenvalue } - \phi \\
\psi_{\lambda} + i \psi_{R \lambda} - \psi_{R^2 \lambda} - i \psi_{R^3 \lambda} & \rightarrow & \textrm{eigenvalue } - i \phi \\
\end{array}
\ed
Note that for weights $\lambda$ with $R^2 \lambda = \lambda$, two of these combinations are zero.

The $\mathbf{6}$ and $\mathbf{10}$ do not posses zero weights, and by checking the eigenvalues we find that the phase $\phi = \pm 1$. The above considerations thus determine the action of $q$ up to an overall sign. In the actual constructions for $E_7$ and $E_8$, this sign will be compensated for by another sign coming from an $SU(2)$-subgroup. 

The $\mathbf{15}$ is the adjoint of $SU(4)$. Among the fifteen weights there are 12 non-zero weights, that decompose in 3 4-cycles under the action of $q$. The remaining three zero weights correspond to the generators of the CSA, the $H_{\alpha}$, on which we already know the action of $q$. The eigenvectors of $q$ are
\bd
\begin{array}{rcl}
H_{\alpha} + H_{R \alpha} + H_{R^2 \alpha} + H_{R^3 \alpha} & = & 0 \\
H_{\alpha} - i H_{R \alpha} - H_{R^2 \alpha} + i H_{R^3 \alpha} & \rightarrow & \textrm{eigenvalue } i \\
H_{\alpha} -  H_{R \alpha} + H_{R^2 \alpha} - H_{R^3 \alpha} & \rightarrow & \textrm{eigenvalue } -1 \\
H_{\alpha} + i H_{R \alpha} - H_{R^2 \alpha} - i H_{R^3 \alpha} & \rightarrow & \textrm{eigenvalue } -i \\
\end{array}
\ed

The last irrep we wish to consider is the $\mathbf{20}$. The easiest way to reconstruct the action of $q$ in this irrep, is to use the decomposition of the tensor product of two $\mathbf{6}$'s: $\mathbf{6 \otimes 6} = \mathbf{1 \oplus 15 \oplus 20}$. From this decomposition we quickly find that the $\mathbf{20}$ has 18 non-zero-weights, which under the action of $q$ decompose in 4 4-cycles and one 2-cycle. The two zero weights form a 2-cycle. 

We note that these considerations imply that
\bd
p T^a p^{-1} = \exp \left( \frac{2 \pi i n}{4} \right) \ T^a, \quad qT^aq^{-1} = \exp \left( \frac{2 \pi i m}{4} \right) \ T^a
\ed

To compute the representations of $SU(4)$ we follow the same route as for $SU(3)$. Define the CSA generators $H_{\alpha'_1}$, $H_{\alpha'_2}$ and $H_{\alpha'_3}$ associated to the simple roots of $su(4)$, following the conventions of the appendix of \cite{Keur2}. Set $H_{\alpha'_i}= h_{\beta_i}$, for $h_{\beta_i} \in {\cal L}_G$. We need the combinations $2 \Lambda'_i / \langle \alpha'_k, \alpha'_k \rangle$ (see eq. (\ref{qk})), where $\Lambda'_i$ are the fundamental weights of $su(4)$ and $\alpha'_k$ are simple roots of $su(4)$. In our conventions $\langle \alpha'_k, \alpha'_k \rangle = 1/4$, for $\alpha'_k$ a root of $su(4)$. The combinations we need are thus 
\bd
 8 \Lambda'_1 = 6 \beta_1 + 4 \beta_2 + 2 \beta_3 \qquad 8 \Lambda'_2 = 4 \beta_1 + 8 \beta_2 + 4 \beta_3 \qquad 8 \Lambda'_3 = 2 \beta_1 + 4 \beta_2 + 6 \beta_3.
\ed 
The $q_k$ from eq. (\ref{qk}) can then be found by computing the inner products of $8 \Lambda'_i$ with the weights of $G$.

\subsection{Realisations of the $SU(4)$-based construction}
\subsubsection{$E_7$}

As mentioned before, we need the subgroup of $E_7$ whose universal cover is  $SU(2) \times SU(4) \times SU(4)$. This subgroup can be chosen to be regular, such that its lattice is a sublattice of the $E_7$-lattice. We take as $SU(2)$-root:
\be
(\sqrt{2},0,0,0,0,0,0)/6
\ee
For the first $SU(4)$ factor that we will label $SU(4)_1$, we take as roots:
\be
\alpha_1^1 = (0,0,0,0,0,1,1)/6 \qquad \alpha_2^1 = (0,0,0,0,1,-1,0)/6 \qquad \alpha_3^1 = (0,0,0,0,0,1,-1)/6
\ee
The second $SU(4)$-factor ($SU(4)_2$) has as roots:
\be
\alpha_1^2 = (0,0,1,1,0,0,0)/6 \qquad \alpha_2^2 = (0,1,-1,0,0,0,0)/6 \qquad \alpha_3^2 = (0,0,1,-1,0,0,0)/6
\ee
With these roots we have the following branching rules for $E_7$ into $SU(2) \times SU(4) \times SU(4)$
\bea
E_7 & \rightarrow & SU(2) \times SU(4) \times SU(4) \non \\
\mathbf{56} & \rightarrow & \mathbf{(1,4,\bar{4}) \oplus (1,\bar{4},4) \oplus (2,6,1) \oplus (2,1,6)} \\
\mathbf{133} & \rightarrow & \mathbf{(1,15,1) \oplus (1,1,15) \oplus (3,1,1) \oplus (1,6,6) \oplus (2,4,4) \oplus (2,\bar{4},\bar{4})} \non
\eea
In the $SU(4) \times SU(4)$-part we construct as usual a diagonal subgroup $D$. We only list the Cartan generators:
\bea
H^D_{\alpha_1} & = & \frac{3}{4}h_{(0,0,1,1,0,1,1)} \non \\
H^D_{\alpha_2} & = & \frac{3}{4}h_{(0,1,-1,0,1,-1,0)} \\
H^D_{\alpha_3} & = & \frac{3}{4}h_{(0,0,1,-1,0,1,-1)} \non
\eea
The $SU(2)$-factor has as its generators
\be
L_3 =  3 h_{(\sqrt{2},0,0,0,0,0,0)} \qquad L_+ = \sqrt{18} e_{(\sqrt{2},0,0,0,0,0,0)}
\ee
We compute the inner products of the weights of $E_7$ with $(3 \sqrt{2},0,0,0,0,0,0)$ to find the $SU(2)$-representations (if the inner product is integer, the corresponding weight belongs to an $SO(3)$ representation, otherwise it is a genuine $SU(2)$ representation with $\mathbf{Z}_2$-centre \cite{Keur2}). To check the $SU(4)$ irreps, it is sufficient to compute $q_1$ (see (\ref{qk})) for the diagonal group. $q_1$ is found by computing the inner product of the weights of $E_7$ with $6 \alpha_1^D + 4 \alpha_2^D + 2 \alpha_3^D = (0,3,3,3,3,3,3)$ (where $\alpha_i^D = \alpha_i^1 + \alpha_i^2$ are the simple roots for the diagonal group). If $q_1 \textrm{ mod }1= 3/4$ the weight belongs to a representation of the fundamental $SU(4)$, for $q_1 \textrm{ mod }1 = 1/4$ it belongs to an irrep that is congruent to the complex conjugate of the fundamental, for $q_1 \textrm{ mod }1 = 1/2$ the irrep is congruent to $SU(4)/\mathbf{Z}_2$, and if $q_1$ is an integer the irrep is congruent to $SU(4)/\mathbf{Z}_4$. Careful examination shows that all weights of $E_7$ fall into two categories: One category is formed by weights that are in an irrep of $(SU(4)/\mathbf{Z}_4) \times SO(3)$, while the second category consist of weights that are in an faithful irrep of $(SU(4)/\mathbf{Z}_2) \times SU(2)$.    
We now construct the elements
\bea
P_{SU(4)} & = & \exp \left( 2 \pi i  H^D_{3 \alpha_1 + 4 \alpha_2 + 3 \alpha_3} \right) \\
Q_{SU(4)} & = & \exp \frac{\pi i}{2}( (1-i)(E^D_{\alpha_1} + E^D_{\alpha_2} + E^D_{\alpha_3} + E^D_{-\alpha_1- \alpha_2 - \alpha_3}) -(E^D_{\alpha_1+ \alpha_2} + E^D_{\alpha_2+ \alpha_3}) \\ & & \qquad -(E^D_{-\alpha_1-\alpha_2} + E^D_{-\alpha_2 - \alpha_3}) + (1+i)(E^D_{-\alpha_1} + E^D_{-\alpha_2} + E^D_{-\alpha_3} + E^D_{\alpha_1 + \alpha_2 + \alpha_3})) \non
\eea
in the diagonal $SU(4)$-group, and
\be
P_{SU(2)} = \exp(i \pi L_3) \qquad Q_{SU(2)} = \exp \left( \frac{i \pi}{2}(L_+ + L_-) \right)
\ee
in the $SU(2)$ group. For a weight $\lambda$ with corresponding eigenvector $\psi_{\lambda}$, that belongs to an $(SU(4)/\mathbf{Z}_4) \times SO(3)$-irrep, we have 
\bd
P_{SU(4)} Q_{SU(4)} \psi_{\lambda} = Q_{SU(4)} P_{SU(4)} \psi_{\lambda} \qquad P_{SU(2)} Q_{SU(2)} \psi_{\lambda} = Q_{SU(2)} P_{SU(2)} \psi_{\lambda}
\ed
If the weight $\lambda$ belongs to a faithful $(SU(4)/\mathbf{Z}_2) \times SU(2)$-irrep, we have 
\bd
P_{SU(4)} Q_{SU(4)} \psi_{\lambda} = -Q_{SU(4)} P_{SU(4)} \psi_{\lambda} \qquad P_{SU(2)} Q_{SU(2)} \psi_{\lambda} = - Q_{SU(2)} P_{SU(2)} \psi_{\lambda}
\ed

The products
\be
P = P_{SU(2)} P_{SU(4)} = \exp (3 \pi i h_{(\sqrt{2},2,1,0,2,1,0)}), \qquad Q = Q_{SU(2)} Q_{SU(4)}
\ee
are commuting elements of $E_7$, since they commute on each weight of any representation. We construct another diagonal subgroup $D'$ in $SU(4) \times SU(4)$ as usual, by applying the rotation 
\be \label{aR}
R: \alpha_1 \rightarrow \alpha_2 \rightarrow \alpha_3 \rightarrow -(\alpha_1 + \alpha_2 + \alpha_3) \rightarrow \alpha_1,
\ee
to the first $SU(4)$ factor and then reconstructing the diagonal subgroup. In this diagonal subgroup we construct an element
\be 
P'_{SU(4)} = \exp(i \pi H^{D'}_{3 \alpha_1 + 4 \alpha_2 + 3 \alpha_3}) 
\ee
The same considerations as above lead to the conclusion that the element 
\be 
P' = P_{SU(2)} P'_{SU(4)} = \exp (3 \pi i h_{(\sqrt{2},0,-1,-2,2,1,0)})
\ee
commutes with $P$ and $Q$, and hence we have three candidate holonomies $\Omega_1 = P$, $\Omega_2 = P'$ and $\Omega_3 = Q$.

Next we calculate the unbroken subgroup. Computing the commutators of the generators of $E_7$ with $P$ and $P'$, we find that only the CSA-generators of $E_7$ commute with both $P$ and $P'$. These CSA generators correspond to zero weights in the representation of the $SU(2) \times SU(4) \times SU(4)$ subgroup, and from previous considerations we know that there is no combination of zero weights of the adjoints of $SU(4)$ or $SU(2)$ that commutes with $Q$. Hence we find that there is no generator that commutes with $P$, $P'$ and $Q$, and the unbroken subgroup is at most discrete.
 
Instead of the rotation $R$ (\ref{aR}), one could also apply the rotations $R^2$ or $R^3$ to construct elements that we will call $P''$ resp. $P'''$. The holonomies $\Omega_1 = P$, $\Omega_2 = P'''$ and $\Omega_3 = Q$ also only commute with a discrete subgroup of $E_7$, and give a non-trivial flat connection inequivalent to the one implied by $\Omega_1 = P$, $\Omega_2 = P'$ and $\Omega_3 = Q$. The non-trivial flat connection implied by $\Omega_1 = P$, $\Omega_2 = P''$ and $\Omega_3 = Q$ gives a bigger unbroken subgroup. The unbroken symmetry group can be calculated to be $U(1)^3$, and this non-trivial flat connection is actually a gauge deformation of the flat connection constructed in \cite{Keur2}, based on twist in $SU(2)$. Therefore the construction based on twist in $SU(2) \times SU(4) \times SU(4)$ gives two new vacua with discrete symmetry group, and a gauge deformation of a vacuum configuration that we have encountered before.

We wish to point out that also the $SU(4)$-based construction in $E_7$ fits into a pattern. We already remarked that the $G_2$ and $SO(7)$ non-trivial flat connection can be described by the 7 triples $(\pm 1, \pm 1, \pm 1)$ with at least one $-1$, and that the $SU(3)$-based construction in $F_4$ can be characterised by the 26 triples (\ref{f4triples}) with at least one element not equal to 1. For the $SU(4)$-based construction, the eigenvalues that appear on the diagonal of the diagonalised holonomies are $i^n$. Consider now the triples $(i^{n_1}, i^{n_2}, i^{n_3})$ with $n_j \in \mathbf{Z}$. There are $4^3 = 64$ distinct triples. Now exclude all triples that are not of order 4, by which we mean that we demand each triple to contain at least one $i$ or $-i$. We are then left with $64 - 2^3 = 56$ triples. $56$ is precisely the dimension of the fundamental irrep of $E_7$, and indeed, constructing the holonomies in this representation and diagonalising, we find that the triples $((\Omega_1)_{ii}, (\Omega_2)_{ii}, (\Omega_3)_{ii})$ are precisely the triples mentioned above.

\subsubsection{$E_8$}

The $SU(2) \times SU(4) \times SU(4)$ construction of $E_7$ can be easily embedded in $E_8$ by using the fact that the $E_7$-lattice is a sublattice of the $E_8$-lattice. More specific, we can take as $SU(2)$-root:
\be
(1,1,0,0,0,0,0,0)/\sqrt{60}
\ee
One $SU(4)$ factor has as roots:
\bea
\alpha_1 & = & (0,0,0,0,0,0,1,1)/\sqrt{60} \non \\
\alpha_2 & = & (0,0,0,0,0,1,-1,0)/\sqrt{60} \\
\alpha_3 & = & (0,0,0,0,0,0,1,-1)/\sqrt{60} \non
\eea
A second $SU(4)$-factor is generated by the roots:
\bea
\alpha_1 & = &(0,0,0,1,1,0,0,0)/\sqrt{60} \non \\
\alpha_2 & = &(0,0,1,-1,0,0,0,0)/\sqrt{60} \\
\alpha_3 & = &(0,0,0,1,-1,0,0,0)/\sqrt{60} \non
\eea
With this information it is trivial to copy the $E_7$ construction. Like in the $E_7$-case, one finds 3 vacua. For two of these, the subgroup unbroken by the $E_7$ holonomies is $SU(2)$. This is to be expected, since $E_8$ decomposes into $E_7 \times SU(2)$:
\bea
E_8 & \rightarrow & E_7 \times SU(2) \non \\
\mathbf{248} & \rightarrow & \mathbf{(133,1) \oplus (56,2) \oplus (1,3)}
\eea
The third vacuum is, as for $E_7$, a gauge deformation of the vacuum based on twist in $SU(2)$ constructed in \cite{Keur2}.

\section{Non-trivial vacua based on twist in $SU(5)$}

\subsection{Twist in $SU(5)$}

We will take the canonical form (in the conventions of the appendices of \cite{Keur2}), where again we will use $H_{\alpha}$ and $E_{\alpha}$ for the subgroup, and $h_{\alpha}$ and $e_{\alpha}$) for the original group.

In $SU(5)$ we look for two matrices satisfying
\be
pq = \exp(\frac{2 \pi i}{5})qp
\ee
We take:
\be 
p = \left( \begin{array}{ccccc} \exp(\frac{4 \pi i}{5}) & 0 & 0 & 0 & 0  \\ 0 & \exp(\frac{2 \pi i}{5}) & 0 & 0 & 0  \\ 0 & 0 & 1 & 0 & 0 \\ 0 & 0 & 0 & \exp(-\frac{2 \pi i}{5}) & 0 \\
0 & 0 & 0 & 0 & \exp(-\frac{4 \pi i}{5}) \end{array} \right), \qquad 
q = \left( \begin{array}{ccccc} 0 & 1 & 0 & 0 & 0  \\ 0 & 0 & 1 & 0 & 0 \\ 0 & 0 & 0 & 1 & 0 \\ 0 & 0 & 0 & 0 & 1 \\ 1 & 0 & 0 & 0 & 0  \end{array} \right)
\ee

In terms of generators this is
\bea
p & = & \exp \left( 2 \pi i  H_{4 \alpha_1 + 6 \alpha_2 + 6 \alpha_3 + 4 \alpha_4} \right) \\
q & = & \exp \Bigl( \frac{2 \pi i}{5} (a (E_{\alpha_1} + E_{\alpha_2} + E_{\alpha_3} + E_{\alpha_4} + E_{-(\alpha_1 + \alpha_2 + \alpha_3 + \alpha_4)}) + \\ & & \qquad b (E_{\alpha_1+ \alpha_2} + E_{\alpha_2+\alpha_3} + E_{\alpha_3+\alpha_4} + E_{-(\alpha_1+ \alpha_2 + \alpha_3)} + E_{-(\alpha_2 + \alpha_3 + \alpha_4)}) \non \\
& & \qquad + \textrm{complex conjugate}) \Bigr) \non
\eea
with 
\bea
a & = & \left[ \exp \left(-\frac{2 \pi i}{5} \right) - \exp \left( \frac{2 \pi i}{5} \right) \right] + 2 \left[ \exp \left( - \frac{4 \pi i}{5} \right)- \exp \left( \frac{4 \pi i}{5} \right) \right] \\
b & = & \left[ \exp\left( -\frac{4 \pi i}{5} \right) - \exp  \left( \frac{4 \pi i}{5} \right) \right] + 2 \left[ \exp \left( \frac{2 \pi i}{5} \right) - \exp \left( -\frac{2 \pi i}{5} \right) \right]  
\eea

The commutation relations of $p$ and $q$ with the group generators are most easily calculated in a specific representation. One finds
\be
\begin{array}{rclcrcl}
p H_{\alpha} & = & H_{\alpha} p, & & q H_{\alpha} & = & H_{R \alpha} q, \\
p E_{\alpha} & = & \exp (2 \pi i \langle \alpha, 4\alpha_1 + 6\alpha_2 +  6 \alpha_3 + 4\alpha_4  \rangle) E_{\alpha} p, & & q E_{\alpha} & = & E_{R \alpha} q.
\end{array}
\ee
The action of the rotation $R$ (which is in this case a genuine rotation) is fully determined by its action on the simple roots.
\be
R : \qquad \alpha_1 \rightarrow \alpha_2 \rightarrow \alpha_3 \rightarrow \alpha_4 \rightarrow -(\alpha_1 + \alpha_2 + \alpha_3 + \alpha_4) \rightarrow \alpha_1
\ee
Again this can be conveniently depicted by a permutation of the roots of the extended Dynkin diagram of $SU(5)$. $R$ is an element of the Weyl group: It is the composition of the Weyl reflection generated by $\alpha_4$, followed by the reflections generated by $\alpha_3$, $\alpha_2$, and $\alpha_1$.

For the calculation of the unbroken subgroup we will only need the action of $q$ on the zero weights of the adjoint. These are the CSA-generators, on which the action of $q$ is easily diagonalised. The only fact we will need is that there is no linear combination of CSA-generators that commutes with $q$.

\subsection{Realisation of the $SU(5)$-based construction}

\subsubsection{$E_8$}

Only $E_8$ allows a suitable $PG'_2$-subgroup with $\widetilde{PG'}_2 = SU(5) \times SU(5)$. We take as root vectors for the first $SU(5)$-subgroup
\bea
\alpha_1^1 = (\hlf,-\hlf,-\hlf,-\hlf,-\hlf,-\hlf,-\hlf,-\hlf)/\sqrt{60}, & & 
\alpha_2^1 = (0,0,0,0,0,0,1,1)/\sqrt{60}, \\
\alpha_3^1 = (0,0,0,0,0,1,-1,0)/\sqrt{60}, & &  \alpha_4^1 = (0,0,0,0,0,0,1,-1)/\sqrt{60}. \non
\eea
while the second $SU(5)$-factor will have roots
\bea
\alpha_1^2 = (0,0,0,1,-1,0,0,0)/\sqrt{60}, & & \alpha_2^2 = (0,0,1,-1,0,0,0,0)/ \sqrt{60}, \\
\alpha_3^2 = (0,1,-1,0,0,0,0,0)/\sqrt{60}, & & \alpha_4^2 = (-1,-1,0,0,0,0,0,0)/ \sqrt{60}.
\eea
The decomposition of $E_8$ into this $SU(5) \times SU(5)$ for the adjoint is given by\footnote{We will label the irrep with Dynkin labels (1000) as $\mathbf{5}$ and the irrep with labels (0010) as $\mathbf{10}$. We define the congruence class of the representation with Dynkin labels $(n_1,n_2,n_3,n_4)$ to be labelled by the integer $n_1 + 2 n_2 + 3 n_3 + 4 n_4 \textrm{ mod } 5$.}
\bea
E_8 & \rightarrow & SU(5) \times SU(5) \non \\
\mathbf{248} & \rightarrow & \mathbf{(24,1) \oplus (1,24) \oplus (10, 5) \oplus (5, \overline{10}) \oplus (\bar{5},10) \oplus (\overline{10},\bar{5})} 
\eea
The congruence classes of the irreducible components of $SU(5) \times SU(5)$ are then${}^1$ $(0,0)$ with multiplicity 2, and $(3,1)$, $(1,2)$, $(4,3)$ and $(2,4)$ all with multiplicity 1. We note that for each congruence class $(a,b)$ we have $a+2b \textrm{ mod } 5 = 0$. Since all $E_8$ representations are isomorphic, this must automatically hold for any $E_8$ representation.

We now construct the elements
\bea
P_i & = & \exp \left( 2 \pi i  H_{4 \alpha_1^i + 6 \alpha_2^i + 6 \alpha_3^i + 4 \alpha_4^i} \right) \\
Q_i & = & \exp \Bigl( \frac{2 \pi i}{5} (a (E_{\alpha_1^i} + E_{\alpha_2^i} + E_{\alpha_3^i} + E_{\alpha_4^i} + E_{-(\alpha_1^i + \alpha_2^i + \alpha_3^i + \alpha_4^i)}) + \\ & & \qquad b (E_{\alpha_1^i+ \alpha_2^i} + E_{\alpha_2^i+\alpha_3^i} + E_{\alpha_3^i+\alpha_4^i} + E_{-(\alpha_1^i+ \alpha_2^i + \alpha_3^i)} + E_{-(\alpha_2^i + \alpha_3^i + \alpha_4^i)}) \non \\
& & \qquad + \textrm{complex conjugate} ) \Bigr) \non
\eea
When acting on an eigenvector $\psi_{\lambda}$, corresponding to a weight $\lambda$ that belongs to an irrep with $SU(5) \times SU(5)$-congruence class labelled by $(n_1, n_2)$, $P_i$ and $Q_i$ obey the following commutation rule
\be
P_i Q_i \psi_{\lambda} = \exp \left( \frac{2 \pi i n_i}{5} \right) Q_i P_i \psi_{\lambda}
\ee
From the above observation on congruence classes it follows that for any weight $\lambda$ of $E_8$ we have 
\be
P_1 Q_1 (P_2)^2 Q_2 \psi_{\lambda} = \exp \left( \frac{2 \pi i(n_1+ 2n_2)}{5} \right) Q_1 P_1 Q_2 (P_2)^2 \psi_{\lambda} = Q_1 P_1 Q_2 (P_2)^2 \psi_{\lambda}
\ee
Hence the elements
\be
P =  P_1 (P_2)^2, \qquad Q = Q_1 Q_2
\ee 
commute. A third commuting element is constructed in the standard way:
\be 
P' = Q_1^{-n} P Q_1^{n} = Q_2^{n} P Q_2^{-n}
\ee
$n$ can range from $1$ to $4$, and therefore there are 4 different flat connections constructed this way. We will work out the case where $n=1$.

In that case we have
\bea
P & = & \exp \left( \frac{2 \pi i}{5} \sqrt{15}h_{(-6,2,-2,-6,-10,4,2,0)} \right) \\
P'& = & \exp \left( \frac{2 \pi i}{5} \sqrt{15}h_{(-2,-2,-6,-10,6,4,2,0)} \right)
\eea
After a somewhat tedious calculation we find that none of the ladder operators $e_{\beta}$ of $E_8$ commutes with both $P$ and $P'$, so we are only left with the 8 Cartan generators. We have to check whether these commute with $Q$. $Q$ has been constructed in such a way that it is an element of the diagonal $SU(5)$-subgroup, so we can apply our previous methods. The elements of the Cartan subalgebra are also the CSA elements for $SU(5) \times SU(5)$, and we know that none of these are invariant under commutation with $Q$ ($Q$ takes the role of $q$ in the diagonal subgroup of $SU(5) \times SU(5)$). Hence no group generator commutes with $P$, $P'$ and $Q$ simultaneously, and the unbroken subgroup is at most discrete.

For the $SU(5)$ based construction, the eigenvalues that appear in the diagonalised holonomies are $\exp \left( \frac{2 \pi i n}{5} \right)$. Consider now the triples
\be 
( \exp(\frac{2 \pi i n_1}{5}),\exp( \frac{2 \pi i n_2}{5}),\exp(\frac{2 \pi i n_3}{5})) \qquad n_i \in \mathbf{Z} 
\ee
If one excludes the triple $(1,1,1)$, there are $5^3 -1 = 124$ distinct triples of this type. Unlike the cases we considered previously, 124 is not the dimension of any representation of $E_8$. The smallest non-trivial representation of $E_8$ is the 248-dimensional adjoint. Constructing the holonomies in the adjoint and diagonalising, one finds that the 248  triples $(P_{ii}, P'_{ii}, Q_{ii})$ consist of precisely two sets of the above 124 triples.

\section{Non-trivial vacua based on twist in $SU(2) \times SU(3)$} \label{finish}

At first sight all possibilities seem to be exhausted, since constructions with $SU(N)$, $N > 5$ are impossible because the exceptional groups are not big enough to contain more than one $SU(6)$-factor. There is however still one more possibility, based on our previous constructions based on $SU(2)$ and $SU(3)$. It is easy to see that $E_8$ allows the $SU(2)$ non-trivial flat connection to be realised simultaneously with the $SU(3)$-flat connection.

\subsection{$E_8$}

According to the decomposition (\ref{e8g2f4}) $E_8$ allows a $G_2 \times F_4$ subgroup (in this case we are dealing with a subgroup that is a genuine product). In \cite{Keur2} we constructed holonomies for a non-trivial flat connection in $G_2$, that we named $P$, $P'$ and $Q$. Here we will rename them to $P_{G_2}$, $P'_{G_2}$ and $Q_{G_2}$ to avoid confusion. $F_4$ allows different types of non-trivial flat connections. We will use the non-trivial flat connections constructed in section \ref{f4nontriv}, and will rename the holonomies $P$, $P'$, $P''$ and $Q$ constructed there to $P_{F_4}$, $P'_{F_4}$, $P''_{F_4}$ and $Q_{F_4}$. Decomposing $E_8$ into its $G_2 \times F_4$ subgroup, elements of this subgroup can be denoted by a pair of elements $(g_{G_2}, g_{F_4})$ with $g_{G_2} \in G_2$ and $g_{F_4} \in F_4$. Therefore $(P_{G_2}, P_{F_4})$ and $(Q_{G_2}, Q_{F_4})$ represent commuting elements of $E_8$. A third commuting element can be constructed by applying the standard procedure of rotating group factors to both elements of the pair to construct the element $(P'_{G_2}, P'_{F_4})$. The rotation can be applied multiple times to construct multiple triples. We have the following possibilities:
\bea
\Omega_1 = (P_{G_2}, P_{F_4}) & \Omega_2 = (P'_{G_2}, P'_{F_4}) & \Omega_3 = (Q_{G_2}, Q_{F_4}) \\ 
\Omega_1 = (P_{G_2}, P_{F_4}) & \Omega_2 = (P_{G_2}, P''_{F_4}) & \Omega_3 = (Q_{G_2}, Q_{F_4}) \label{su3type1} \\
\Omega_1 = (P_{G_2}, P_{F_4}) & \Omega_2 = (P'_{G_2}, P_{F_4}) & \Omega_3 = (Q_{G_2}, Q_{F_4}) \label{su2type} \\
\Omega_1 = (P_{G_2}, P_{F_4}) & \Omega_2 = (P_{G_2}, P'_{F_4}) & \Omega_3 = (Q_{G_2}, Q_{F_4}) \label{su3type2} \\
\Omega_1 = (P_{G_2}, P_{F_4}) & \Omega_2 = (P'_{G_2}, P''_{F_4}) & \Omega_3 = (Q_{G_2}, Q_{F_4})
\eea
However, not all of these are new. The flat connection implied by the holonomies (\ref{su3type1}) has twice the element $P_{G_2}$ in the $G_2$-factor: This means that in the $G_2$ subgroup the connection can be deformed to a trivial one, and only the $F_4$ part is non-trivial. This is a deformation of one of the flat connections based on $SU(3)$-twist that was already constructed previously. Similarly the flat connection implied by the holonomies (\ref{su3type2}) is deformable to the other flat connection based on $SU(3)$-twist. The flat connection implied by (\ref{su2type}) is trivial in its $F_4$ factor, and only non-trivial in its $G_2$ part. It is therefore a deformation of the flat connection in $E_8$ that was constructed in \cite{Keur2}.
The remaining two flat connections are new. We note that, to commute a generator with the $\Omega_i$, we can commute it first with the $G_2$-element and then with the $F_4$-element. We can choose a basis of generators such that they commute or anti-commute with the $G_2$-elements, and they commute up to a $\mathbf{Z}_3$-element with the $F_4$-elements. We conclude that to commute with the $\Omega_i$ a generator has to commute with both the $G_2$ and the $F_4$-elements. The generators that commute with the $G_2$ part of the holonomies generate the $F_4$ subgroup in the decomposition $G_2 \times F_4$. The $F_4$ part of the holonomies breaks this $F_4$ group completely. Hence only discrete symmetries commute with the holonomies constructed here.

\section{Re-evaluation of $\textrm{Tr}(-)^F$} \label{ind}

One of the motivations for this research was to check whether non-trivial flat connections may solve the problem of calculating $\textrm{Tr}(-)^F$ in supersymmetric gauge theories with exceptional gauge groups. We announced in \cite{Keur2} that it does, and we will now justify this claim. 

We have shown that the bosonic equations of motion allow several solutions, to be divided in several components of the moduli space. Upon quantisation one finds that each component implies a bosonic vacuum \cite{IWit, Witnew}. To perform the vacuum count one has to count in how many ways zero-energy fermions can be added. The reasoning of \cite{IWit} implies that in each case there are $r$ ways to add fermion pairs to the vacuum, with $r$ the rank of the subgroup that is unbroken on the specific component of moduli space. This gives $r$ bosonic vacua, which together with the original vacuum based on a configuration with bosonic fields only gives a contribution of $r+1$ to $\textrm{Tr}(-)^F$. We now summarise the results of this paper and \cite{Keur2}:

\begin{itemize}
\item{$F_4$:}
We have found 4 components of the moduli space of $F_4$:
\begin{itemize}
\item As always there is the component of the trivial flat connections. At a generic point there is a rank 4 unbroken subgroup, giving a contribution of $r+1 = 5$ to $\textrm{Tr}(-)^F$.
\item In \cite{Keur2} we constructed a new component in moduli space that has maximal unbroken symmetry group $SO(3)$. By deforming around this solution one finds a component of moduli space that has a rank 1 unbroken subgroup, giving a contribution of $r+1 = 2$ to $\textrm{Tr}(-)^F$.
\item As announced in \cite{Keur2}, and explicitly constructed here, there are two components in moduli space that have unbroken discrete symmetry group. These solutions cannot be deformed, and give contributions of $r+1 = 1$ to $\textrm{Tr}(-)^F$.
\end{itemize}
The Witten index count for $F_4$ SYM on $T^3 \times R$ thus gives 
\bd
\textrm{Tr}(-)^F = 5 + 2 + (1+1) = 9,
\ed
in full agreement with the infinite volume calculation since the dual Coxeter number $h = 9$ for $F_4$
\item{$E_6$:}
We have found 4 components of the moduli space of $E_6$:
\begin{itemize}
\item The component of the trivial flat connections. At a generic point there is a rank 6 subgroup unbroken, giving a contribution of $r+1 = 7$ to $\textrm{Tr}(-)^F$.
\item In \cite{Keur2} we constructed a point in moduli space that has maximal unbroken symmetry group $SU(3)$. By deforming around this solution one finds a component of moduli space that has a rank 2 subgroup unbroken, giving a contribution of $r+1 = 3$ to $\textrm{Tr}(-)^F$.
\item In this article we explicitly constructed two components in moduli space that have unbroken discrete symmetry group. These solutions cannot be deformed, and each contribute $r+1 = 1$ to $\textrm{Tr}(-)^F$.
\end{itemize}
The Witten index count for $E_6$ SYM on $T^3 \times R$ thus gives 
\bd
\textrm{Tr}(-)^F = 7 + 3 + (1+1) = 12,
\ed
in full agreement with the infinite volume calculation since the dual Coxeter number $h = 12$ for $E_6$.
\item{$E_7$:}
We have found 6 components of the moduli space of $E_7$:
\begin{itemize}
\item The component of the trivial flat connections. At a generic point there is a rank 7 subgroup unbroken, giving a contribution of $r+1 = 8$ to $\textrm{Tr}(-)^F$.
\item In \cite{Keur2} we constructed a point in moduli space that has maximal unbroken symmetry group $Sp(3)$. By deforming around this solution one finds a component of moduli space that has a rank 3 subgroup unbroken, giving a contribution of $r+1 = 4$ to $\textrm{Tr}(-)^F$.
\item In this article we explicitly constructed, based on twist in $SU(3)$-subgroups, two components in moduli space with maximal unbroken symmetry group $SU(2)$. Deformations leave  a rank 1 unbroken subgroup. We get contributions of each $r+1 = 2$ to $\textrm{Tr}(-)^F$.
\item Based on twist in the $SU(4)\times SU(4) \times SU(2)$-subgroups, two more components were constructed that consist of a point in moduli space that has unbroken discrete symmetry group. These solutions cannot be deformed, and each contributes $r+1 = 1$ to $\textrm{Tr}(-)^F$.
\end{itemize}
The Witten index count for $E_7$ SYM on $T^3 \times R$ thus gives  
\bd
\textrm{Tr}(-)^F = 8 + 4 + (2+2) + (1+1) = 18,
\ed
in full agreement with the infinite volume calculation since the dual Coxeter number $h = 18$ for $E_7$.
\item{$E_8$:}
We have found 12 components of the moduli space of $E_8$:
\begin{itemize}
\item The component of the trivial flat connections. At a generic point there is a rank 8 subgroup unbroken, giving a contribution of $r+1 = 9$ to $\textrm{Tr}(-)^F$.
\item In \cite{Keur2} we constructed a point in moduli space that has unbroken symmetry group $F_4$. By deforming around this solution one finds a component of moduli space that has a rank 4 subgroup unbroken, giving a contribution of $r+1 = 5$ to $\textrm{Tr}(-)^F$.
\item In this article, based on twist in $SU(3)$-subgroups, two components in moduli space were constructed with maximal unbroken symmetry group $G_2$. At a generic point of each of the two components there is a rank 2 subgroup unbroken. Each component contributes $r+1 = 3$ to $\textrm{Tr}(-)^F$.
\item Based on twist in the $SU(4)\times SU(4) \times SU(2)$-subgroups, two more components in moduli space were found with maximal unbroken $SU(2)$ symmetry group. Deformations always leave a rank 1 subgroup unbroken. The components contribute each $r+1 = 2$ to $\textrm{Tr}(-)^F$.
\item With twist in the $SU(5) \times SU(5)$-subgroup, four more components were constructed with unbroken discrete symmetry group. These solutions cannot be deformed, and contribute each $r+1 = 1$ to $\textrm{Tr}(-)^F$.
\item Last, by using $(SU(2)\times SU(3))^2$-subgroups, two more components were constructed that have unbroken discrete symmetry group. These solutions can also not be deformed, and contribute each $r+1 = 1$ to $\textrm{Tr}(-)^F$.
\end{itemize}
The Witten index count for $E_8$ SYM on $T^3 \times R$ thus gives  
\bd
\textrm{Tr}(-)^F =  9 + 5 + (3+3) + (2+2) + (1+1+1+1) + (1+1) = 30
\ed
in agreement with the infinite volume calculation since the dual Coxeter number $h = 30$ for $E_8$.
\end{itemize}

\section{Discussion and conclusions}

The construction of non-trivial flat connections with our method reveals that the moduli spaces of exceptional gauge theories on $T^3 \times R$ consists of many components. The Witten index count indicates that all non-trivial flat connections have been found. Thus for supersymmetric Yang-Mills theory both the two extreme cases of the infinite volume ($R^4$) and small volume ($T^3 \times R$) have $h$ vacua, with $h$ the dual Coxeter number of the gauge group. It would be interesting to learn more about the intermediate volume regime, and the character of the $h$ vacua there. More open questions are the Chern-Simons functional of the gauge field configuration, and the chirality assignment of the different disconnected components.  

An intriguing question is what happens for higher dimensional tori. It is clear that the non-trivial flat connections for the 3--torus can always be trivially embedded on an $n$--torus, by taking $(n-3)$ holonomies equal to the identity, and the remaining 3 holonomies given by our construction. However, also more general configurations are possible. Our method is basically a construction of commuting elements for certain gauge groups, and a set of $n$ commuting elements may be used as holonomies on an $n$--torus. For $SU(N)$ and $Sp(N)$ theories the moduli space always consists of one component, regardless of $n$, but for other groups the number of components of moduli space might grow further with $n$.

We expect that the flat connections constructed in this paper are of relevance in toroidal compactifications of the $E_8 \times E_8$ heterotic string, just like the original constructions of \cite{Witnew} are of relevance for the string theories with $SO(32)$ gauge groups. $E_8$ has a rich moduli space of 12 components, but our analysis shows that many components survive for exceptional subgroups of $E_8$. Another question is what happens to all these vacuum solutions under various dualities. We hope that explicit expressions for the holonomies, as we have constructed, may be helpful in answering these questions.

\acknowledgments

We would like to thank Pierre van Baal, Jan de Boer, Robbert Dijkgraaf, Arkady Vainshtein and Andrei Smilga for helpful discussions.


\begin{thebibliography}{99}
\bibitem{IWit} E. Witten, \npb{202}{1982}{253}.
\bibitem{IDyn} S.F. Cordes and M. Dine, \npb{273}{1986}{581}; A. Morozov, M. Ol'shanetsky and M. Shifman, \npb{304}{1988}{291}.
\bibitem{Witnew} E. Witten, \jhep{02}{1998}{006}, \hepth{9712028}.
\bibitem{Keur1} A. Keurentjes, A. Rosly and A.V. Smilga, \prd{58}{1998}{081701},\\ \hepth{9805183}
\bibitem{Keur2} A. Keurentjes, {\it Non-trivial flat connections on the 3-torus: $G_2$ and the orthogonal groups}, \hepth{9901154}.
\bibitem{Hooft} G.'t Hooft, \npb{153}{1979}{141}.
\bibitem{Cohen} E. Cohen and C. Gomez, \npb{223}{1983}{183}.
\bibitem{Dynkin} E.B. Dynkin, {\it Mat. Sbornik} {\bf 30} ({\bf 72}) 349 (English translation: {\it Amer. Math. Soc. Trans. Series 2}, {\bf 6}, 111).
\bibitem{McKay} W.G. McKay and J. Patera, {\it Tables of dimensions, indices, and branching rules for representations of simple Lie algebra's}  (Marcel Dekker, New York and Basel 1981).
\bibitem{Kac} V.G. Kac and A.V. Smilga, {\it Vacuum structure in supersymmetric Yang-Mills theories with any gauge group}, \hepth{9902029}
\end{thebibliography}
\end{document}